\documentclass[12pt,preprint]{aastex}
\usepackage{apjfonts}
\usepackage{amsmath,amsfonts,amsthm,bm} 
\usepackage{lscape}

\newcommand{\pivec}{\mbox{\boldmath $\pi$}}
\newcommand{\muvec}{\mbox{\boldmath $\mu$}}

\lefthead{JUNG ET AL.}
\righthead{ }

\begin{document}
\title{KMT-2017-BLG-0165Lb: A Super-Neptune mass planet Orbiting a Sun-like Host Star}

\author{
Youn~Kil~Jung$^{1,16}$, 
Andrew~Gould$^{1,2,3,16}$,
Weicheng~Zang$^{4,16,17}$,
Kyu-Ha~Hwang$^{1,16}$,
Yoon-Hyun~Ryu$^{1,16}$, 
Cheongho~Han$^{5,16}$,
Jennifer~C.~Yee$^{6,16}$ \\
and \\
Michael~D.~Albrow$^{7}$, Sun-Ju~Chung$^{1,8}$, In-Gu~Shin$^{6}$, Yossi~Shvartzvald$^{9}$, Wei~Zhu$^{10}$,  
Sang-Mok~Cha$^{1,11}$, Dong-Jin~Kim$^{1}$, Hyoun-Woo~Kim$^{1}$, Seung-Lee~Kim$^{1,8}$, 
Chung-Uk~Lee$^{1,8}$, Dong-Joo~Lee$^{1}$, Yongseok~Lee$^{1,11}$, Byeong-Gon~Park$^{1,8}$, Richard~W.~Pogge$^{2}$ \\
(The KMTNet Collaboration) \\
Matthew~T.~Penny$^{2}$, Shude~Mao$^{4,12,13}$, Pascal~Fouqu\'e$^{14,15}$, Tianshu~Wang$^{4}$ \\
(The CFHT Collaboration) \\
}

\bigskip\bigskip
\affil{$^{1}$Korea Astronomy and Space Science Institute, Daejon 34055, Republic of Korea}
\affil{$^{2}$Department of Astronomy, Ohio State University, 140 W. 18th Ave., Columbus, OH 43210, USA}
\affil{$^{3}$Max-Planck-Institute for Astronomy, K$\rm \ddot{o}$nigstuhl 17, 69117 Heidelberg, Germany}
\affil{$^{4}$Physics Department and Tsinghua Centre for Astrophysics, Tsinghua University, Beijing 100084, China}
\affil{$^{5}$Department of Physics, Chungbuk National University, Cheongju 28644, Republic of Korea}
\affil{$^{6}$Harvard-Smithsonian Center for Astrophysics, 60 Garden St., Cambridge, MA 02138, USA}
\affil{$^{7}$University of Canterbury, Department of Physics and Astronomy, Private Bag 4800, Christchurch 8020, New Zealand}
\affil{$^{8}$Korea University of Science and Technology, 217 Gajeong-ro, Yuseong-gu, Daejeon 34113, Korea}
\affil{$^{9}$IPAC, Mail Code 100-22, Caltech, 1200 E. California Blvd., Pasadena, CA 91125, USA}
\affil{$^{10}$Canadian Institute for Theoretical Astrophysics, University of Toronto, 60 St George Street, Toronto, ON M5S 3H8, Canada}
\affil{$^{11}$School of Space Research, Kyung Hee University, Yongin 17104, Republic of Korea}
\affil{$^{12}$National Astronomical Observatories, Chinese Academy of Sciences, A20 Datun Rd., Chaoyang District, Beijing 100012, China}
\affil{$^{13}$Jodrell Bank Centre for Astrophysics, Alan Turing Building, University of Manchester, Manchester M13 9PL, UK}
\affil{$^{14}$CFHT Corporation, 65-1238 Mamalahoa Hwy, Kamuela, Hawaii 96743, USA}
\affil{$^{15}$Universit\'e de Toulouse, UPS-OMP, IRAP, Toulouse, France}
\footnotetext[16]{The KMTNet Collaboration.}
\footnotetext[17]{The CFHT Collaboration.}

\begin{abstract}
We report the discovery of a low mass-ratio planet $(q = 1.3\times10^{-4})$, 
i.e., 2.5 times higher than the Neptune/Sun ratio. The planetary system was 
discovered from the analysis of the KMT-2017-BLG-0165 microlensing event, 
which has an obvious short-term deviation from the underlying light curve 
produced by the host of the planet. Although the fit improvement with the 
microlens parallax effect is relatively low, one component of the parallax 
vector is strongly constrained from the light curve, making it possible to 
narrow down the uncertainties of the lens physical properties. A Bayesian 
analysis yields that the planet has a super-Neptune mass $(M_{2}=34_{-12}^{+15}~M_{\oplus})$ 
orbiting a Sun-like star $(M_{1}=0.76_{-0.27}^{+0.34}~M_{\odot})$ located at $4.5~{\rm kpc}$. 
The blended light is consistent with these host properties. The projected planet-host separation 
is $a_{\bot}={3.45_{-0.95}^{+0.98}}~{\rm AU}$, implying that the planet is located outside the 
snowline of the host, i.e., $a_{sl}\sim2.1~{\rm AU}$. KMT-2017-BLG-0165Lb is the sixteenth 
microlensing planet with mass ratio $q<3\times10^{-4}$. Using the fifteen of these planets 
with unambiguous mass-ratio measurements, we apply a likelihood analysis to investigate the 
form of the mass-ratio function in this regime. If we adopt a broken power law for the form 
of this function, then the break is at $q_{\rm br}\simeq0.55\times10^{-4}$, which is much 
lower than previously estimated. Moreover, the change of the power law slope, $\zeta>3.3$ 
is quite severe. Alternatively, the distribution is also suggestive of a ``pile-up'' of planets 
at Neptune-like mass ratios, below which there is a dramatic drop in frequency.
\end{abstract}

\section{Introduction}

Up to now, there have been many discoveries of planets via various detection methods, 
which are currently reaching about 4000 planets according to the NASA  Exoplanet Archive
\footnote{https://exoplanetarchive.ipac.caltech.edu}. Most of these planets were discovered 
and characterized by the transit (e.g., \citealt{tenenbaum14}) and radial-velocity 
(e.g., \citealt{pepe11}) methods. These methods favor the detection of close-in planets 
around their hosts stars, and the majority of the hosts are Sun-like stars located 
in the solar neighborhood.

On the other hand, the microlensing method favors the detection of planets down to 
one Earth mass orbiting outside the snow line, where the temperature is cool enough 
for icy material to condense \citep{ida04}. Instead of looking for light from 
the host stars, microlensing uses the light of a background source refracted by the 
gravitational potential of an aligned foreground planetary system. This allows 
the method to detect planets around all types of stellar objects at Galactocentric 
distances and even free-floating planets, which may have been ejected from their host 
stars \citep{sumi11,mroz17b,mroz18}. Hence, although the number of planets discovered 
by microlensing is relatively small ($\sim 60$ discoveries to date), the method can 
access a class of planets that are inaccessible to other detection methods, and help 
to expand our understanding of planet populations.

Based on microlensing planets, statistical works have been conducted to examine 
the properties of planets beyond the snow line. As presented in \citet{mroz17a}, 
microlensing planets are nearly uniformly distributed in ${\rm log}~q$, where 
$q = M_{2}/M_{1}$ is the planet/host mass ratio. Because the probability for 
detecting planets increases with $q$, this implies that the planet frequency 
has a rising shape toward lower mass ratios. \citet{sumi10} investigated this 
distribution and found that the frequency can be described by a single power-law 
mass-ratio function, i.e., $dN/d{\rm log}~q \propto q^{n}$, where $n$ is the power-law 
index. However, \citet{suzuki16} argued that there exists a break in the power-law 
function at $q_{\rm br} \simeq 1.7\times10^{-4}$. They found that the planet frequency 
rises rapidly toward lower mass ratios above the break, but falls off just as rapidly 
toward lower mass ratios below the break. \citet{udalski18} confirmed this turnover 
by refining the power-law index using seven microlensing planets in the $q < 10^{-4}$ 
regime. This broken power law would imply that the planet/host mass ratio provides 
a strong constraint to planet formation beyond the snow line. Furthermore, considering 
that the mass ratio at the break corresponds to $\sim 20~M_{\oplus}$ for the median host 
mass of $0.6~M_{\odot}$, it also would suggest that Neptune-mass planets are the most 
common population of planets outside the snow line \citep{gould06}. However, the precise 
mass ratio at the break remains uncertain, and the nature of the rapid descent below the 
break is barely probed due to the insufficient sample in this regime. Hence, discovering 
microlensing planets whose mass ratios are located near or below the break is important 
to improve our understanding of planet abundances.

In this paper, we report the discovery of a super-Neptune mass planet hosted 
by a Sun-like star, i.e., with a mass ratio close to $q_{\rm br}$ as found by \citet{suzuki16}. 
The planetary system was identified from the analysis of microlensing event KMT-2017-BLG-0165, 
which has an obvious short-term deviation from its underlying single-lens light curve \citep{paczynski86}. 
Despite the short duration, the deviation was clearly detected from the Korea Microlensing Telescope 
Network \citep[KMTNet:][]{kim16} high-cadence microlensing survey. We also investigate the value of 
$q_{\rm br}$ in more detail.

\section{Observation}

The lensing event KMT-2017-BLG-0165 occurred on a star located at 
$({\rm RA}, {\rm Dec})_{\rm J2000} = $(17:58:35.92, $-28$:08:01.21) or  
$(l,b) = (2.14, -2.04)$ in Galactic coordinates. It was found by applying 
the KMTNet event-finder algorithm \citep{kim18} to the 2017 KMTNet survey data 
from three 1.6 m telescopes distributed over three different sites, i.e., 
the Cerro Tololo Inter-American Observatory in Chile (KMTC), 
the South African Astronomical Observatory in South Africa (KMTS), and 
the Siding Spring Observatory in Australia (KMTA). 
The event is located in four overlapping fields (BLG02, BLG03, BLG42, and BLG43), 
monitored with a combined cadence of $8~{\rm hr}^{-1}$. KMTNet images were obtained 
primarily in $I$ band, while some $V$-band images were obtained solely to secure the 
colors of the source stars. All data for the light curve analysis were reduced using 
the pySIS method \citep{albrow09}, a variant of difference image analysis 
\citep[DIA:][]{alard98}.

\begin{figure}[th]
\epsscale{0.9}
\plotone{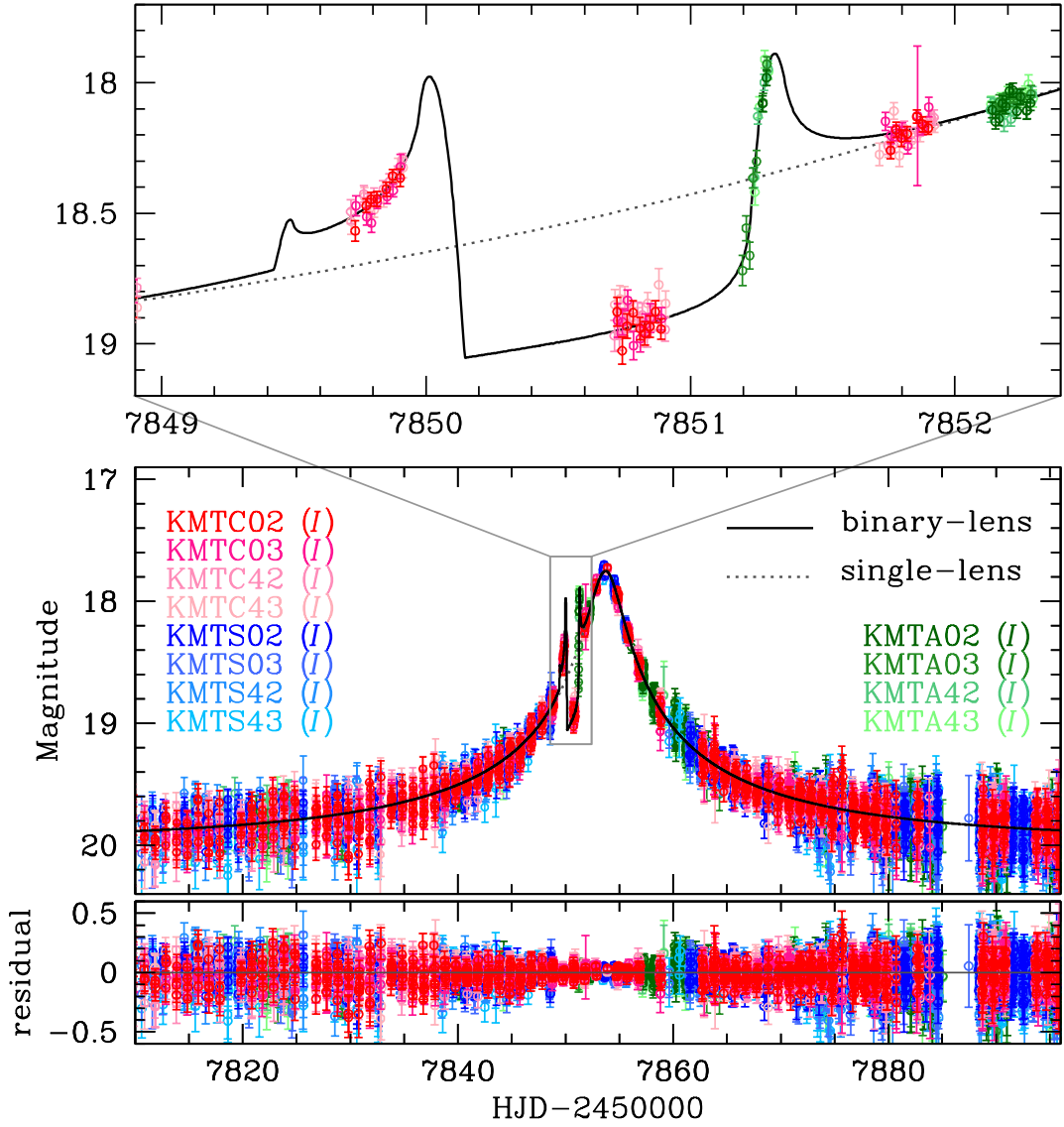}
\caption{\label{fig:one}
Light curve of KMT-2017-BLG-0165. The upper panel shows the 
zoom of the planetary perturbation centered at ${\rm HJD}' \sim 7850.6$. 
The black curve is the model derived from the binary-lens 
interpretation, whereas the dotted grey curve is derived from the single-lens interpretation.
}
\end{figure}

Figure~\ref{fig:one} shows the light curve of KMT-2017-BLG-0165. 
Except for a deviation during the interval $7849 < {\rm HJD}'(={\rm HJD}-2450000) < 7852$, 
the overall shape of the light curve resembles a standard single-lens curve with a 
moderate magnification $A_{\rm max} \sim 32$ at the peak. The most prominent feature of 
the deviation is a trough lying about 0.4 mag below the level of the single-lens curve. 
This trough is a characteristic feature of lensing systems with planet-host separations 
$s < 1$ (in units of the angular Einstein radius, $\theta_{\rm E}$) and small mass ratios 
$q \ll 1$, i.e., planetary systems. In the underlying microlensing event, the host of a 
planet will split the source light into two magnified images, one inside (minor) and 
the other outside (major) the Einstein ring. Because the former image is highly unstable, 
it will be easily suppressed if the planet lies in or near the path of the image, thereby 
causing a trough in the light curve \citep{gaudi12}. Such troughs are always flanked 
by two triangular caustics centered on the host-planet axis. If the source passes very 
close to or over these caustics, the light curve would exhibit sharp breaks near the trough. 
Hence, from the form of the deviation at ${\rm HJD}' \sim 7851.2$, it is suggested 
that there exists an interaction between the source and the triangular caustics.      

\section{Analysis}

Considering that the light curve appears to be a planetary event, we model the event with 
the binary-lens interpretation. For a standard binary-lens model, one needs seven fitting 
parameters. Three of these parameters $(t_{0}, u_{0}, t_{\rm E})$ describe the source approach 
relative to the lens, and their definitions are the same as those for the \citet{paczynski86} curve, 
i.e., the moment of closet approach, the impact parameter (in unit of $\theta_{\rm E}$), and 
the Einstein timescale. Another three parameters $(s, q, \alpha)$ describe the binary lens 
companion. As already described, $s$ and $q$ are the normalized separation and the mass ratio, 
whereas $\alpha$ is the angle of the source trajectory relative to the binary axis. 
The last parameter, $\rho_{*}$, is the source radius $\theta_{*}$ in units of $\theta_{\rm E}$.

For the case that planetary deviations can be regarded as perturbations, the light curve 
except the deviations follows the underlying standard single-lens curve. The fitting parameters 
can then be estimated heuristically from the location and duration 
of the deviation \citep{gould92b}. By excluding the perturbation in the light curve, we find that 
the single-lens fit yields 
$(t_{0}, u_{0}, t_{\rm E}) = (7853.70, 0.035, 42.03~{\rm days})$, which corresponds to an 
effective timescale of $t_{\rm eff} \equiv u_{0}t_{\rm E} = 1.47~{\rm days}$. The perturbation 
is centered at ${\delta}t = 3.2$ days before the peak, implying that the trajectory angle is 
$\alpha = {\rm tan}^{-1}(t_{\rm eff}/{\delta}t) = 5.85$ rad. The normalized separation $s$ is 
then estimated by 
\begin{equation}
s - {1 \over s} = u_{0}\sqrt{1 + \left({{\delta}t \over t_{\rm eff}}\right)^2}, 
\label{eq1}
\end{equation}
which indicates $s = 0.96$. Note that we exclude the other $(s > 1)$ solution because it would 
produce a positive deviation. The mass ratio $q$ is related to the duration of the trough ${\Delta}t$ 
and the separation by
\begin{equation}
q = {\eta_{c,-}^{2} \over 4(s^2 - 1)};~~~~~{\Delta}t = 2t_{\rm E}\eta_{c,-}|{\rm sec}~\alpha|, 
\label{eq2}
\end{equation}
where $\eta_{c,-}$ is the vertical position of triangular-caustic-fold facing the trough \citep{han06}. 
From the light curve, we estimate ${\Delta}t \sim 0.8~{\rm days}$. 
Hence, the mass ratio is $q \sim 2.2\times10^{-4}$.

\begin{figure}[th]
\epsscale{0.9}
\plotone{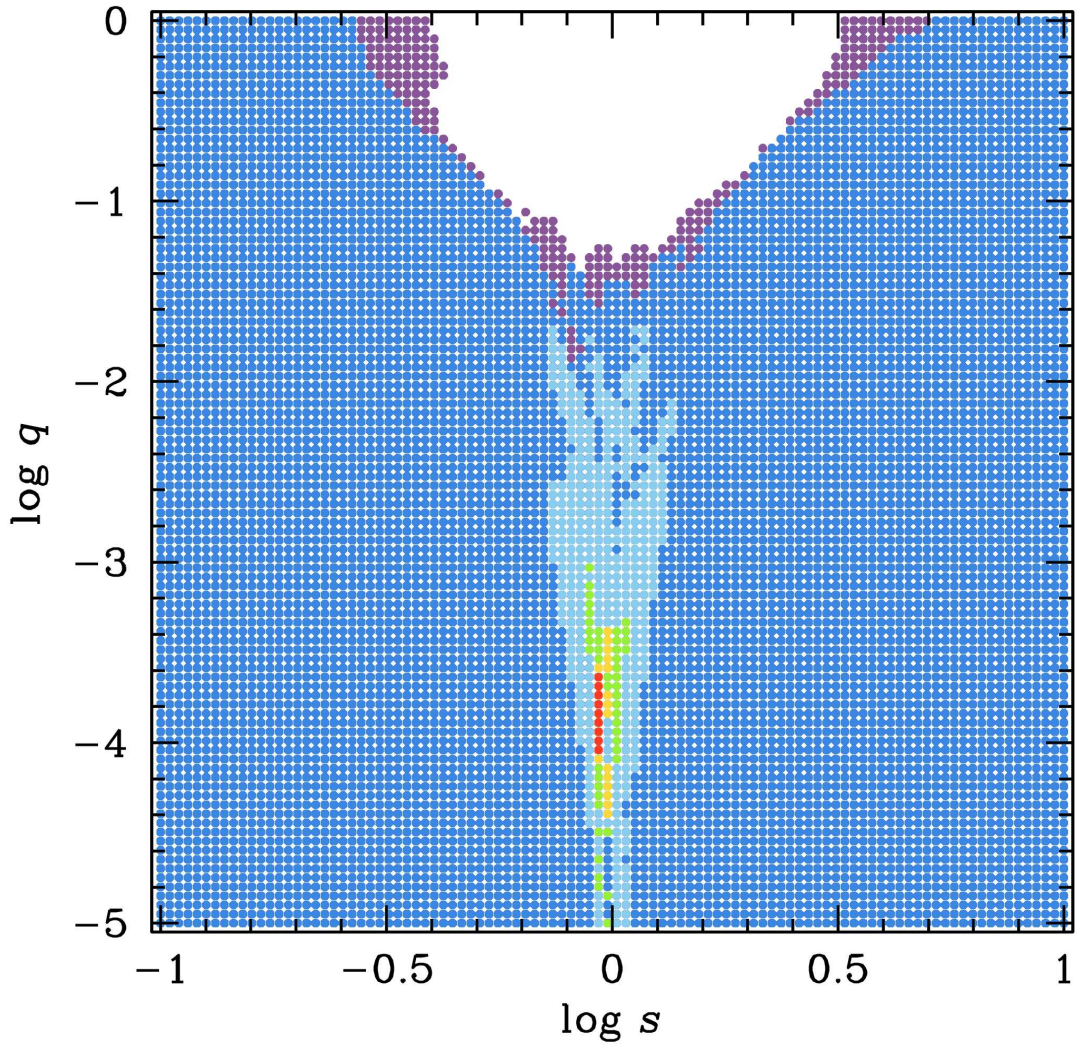}
\caption{\label{fig:two}
$\Delta\chi^{2}$ surface in $({\rm log}~s, {\rm log}~q)$ plane drawn from the grid search. 
The plane is color coded by $\Delta\chi^2$ $< (1\,n)^{2}$ (red), $< (2\,n)^{2}$ (yellow), 
$< (3\,n)^{2}$ (green), $< (4\,n)^{2}$ (light blue), $< (5\,n)^{2}$ (blue), 
and $< (6\,n)^{2}$ (purple) level from the initial best-fit solution, where $n = 20$. 
}
\end{figure}

\begin{figure}[th]
\epsscale{0.9}
\plotone{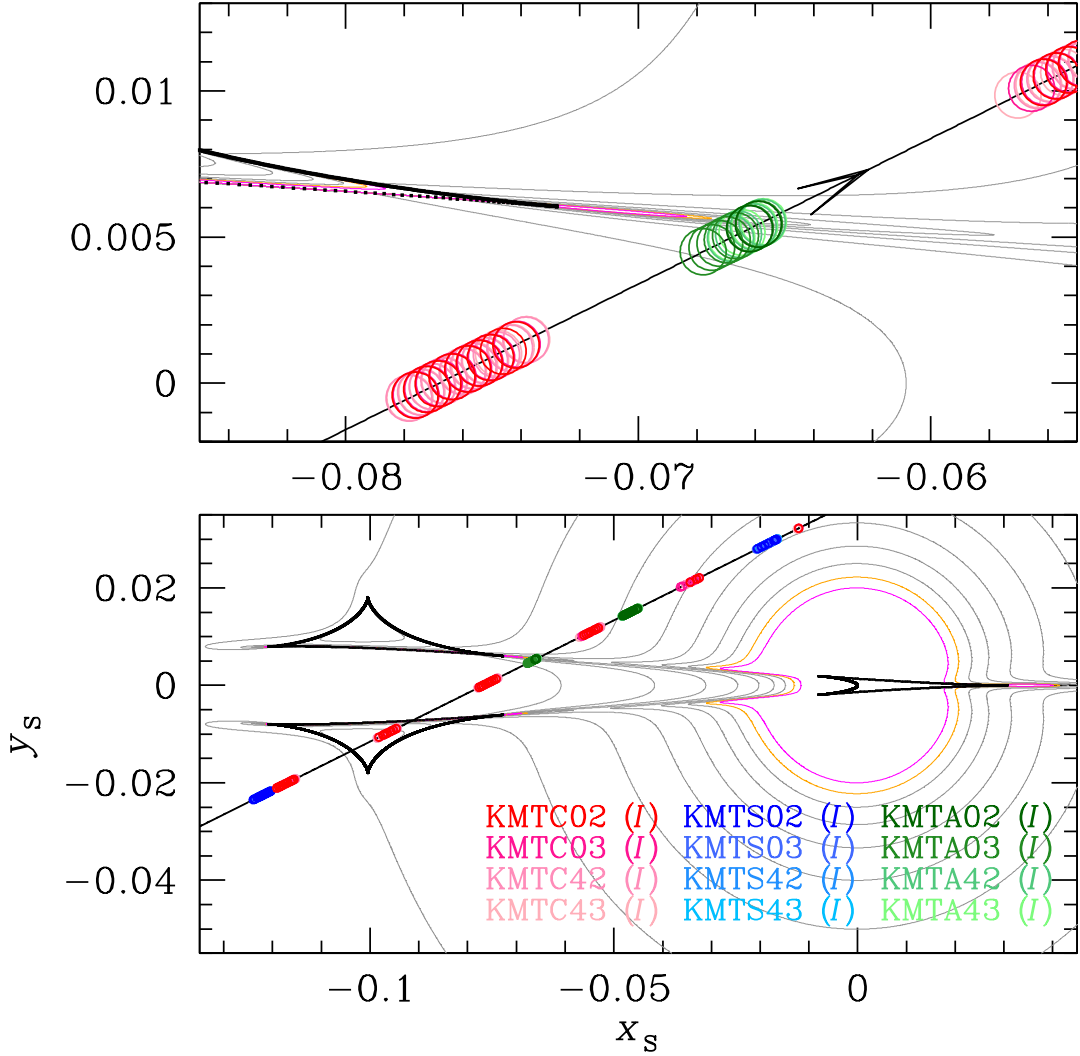}
\caption{\label{fig:three}
Caustic structure of KMT-2017-BLG-0165. The source passes over the trough 
flanked by the two triangular caustics, resulting from the minor image perturbation. 
The upper panel shows the zoom of the triangular caustic at the time of the source's cusp approach. 
The open circles represent the source location at the times of observation, 
and their size is scaled by the source radius $\rho_{*}$ of the best-fit solution. 
The grey curves are the magnification contours of $A = (10, 15, 20, 25, 30, 35, 40)$, 
while the yellow and magenta curves are the contours of $A = 45$ and $50$, respectively.
}
\end{figure}

For precise measurements of the fitting parameters, we conduct a systematic analysis 
by adopting the method of \citet{jung15}. First, we perform a grid search over $(s, q, \alpha)$ 
space, in which we fit the light curve using a downhill approach with a Markov Chain Monte 
Carlo (MCMC) algorithm. At each data point, we compute the magnification using inverse 
ray-shooting \citep{kayser86,schneider87} in the neighboring region around caustics and 
the semi-analytic multipole approximation \citep{pejcha09,gould08} elsewhere. Next, 
we investigate the local $\chi^2$ minima from the derived $\Delta\chi^2$ map 
in the $({\rm log}~s,~{\rm log}~q)$ space. From this, we find only one local minimum, 
which has a separation that is slightly lower than unity and a mass ratio that is in the 
range of $0.8\times10^{-4} < q < 3\times10^{-4}$. These are very close to the predictions 
from the heuristic analysis (see Figure~\ref{fig:two}). To find the global solution, 
we then explore the local solution by optimizing all fitting parameters. The best-fit 
standard parameters are given in Table~\ref{table:one}. 
In Figure~\ref{fig:one}, we present the model curve superposed on the data. 
The corresponding caustic structure is shown in Figure~\ref{fig:three}.

For some binary-lens events, the observed light curve can exhibit further 
deviations from the form expected from the standard model. This often occurs 
in long-timescale events for which the approximation that the source motion 
relative to the lens is rectilinear is no longer valid. There are two major 
effects that can cause such deviations, i.e., ``microlens-parallax'' and 
``lens-orbital'' effects. The former is caused by the orbital acceleration 
of Earth \citep{gould92a}, while the latter is caused by the orbital 
acceleration of the lens \citep{dominik98,jung13}. The parallax effect is 
described by two parameters $\pivec_{\rm E} = (\pi_{{\rm E},N}, \pi_{{\rm E},E})$, 
which are the components of parallax vector, i.e.,
\begin{equation}
\pivec_{\rm E} \equiv \pi_{\rm E}{\muvec_{\rm geo} \over \mu_{\rm geo}};~~~~~\pi_{\rm E} = {\pi_{\rm rel} \over \theta_{\rm E}},
\label{eq3}
\end{equation}
where $\pi_{\rm rel} = {\rm AU}(D_{\rm L}^{-1} - D_{\rm S}^{-1})$ is the 
relative lens-source parallax, and $D_{\rm L}$ and $D_{\rm S}$ are, respectively, 
the lens and source distances. Here $\muvec_{\rm geo}$ is the geocentric lens-source 
relative proper motion. The lens-orbital effect is described by two linearized parameters 
$\bm{\gamma} = [(ds/dt)/s, d\alpha/dt]$, which denote the instantaneous time derivatives 
of $\ln s$ and $\alpha$, respectively. These orbit parameters can be correlated with 
$\pivec_{\rm E}$, and thus they also should be included when incorporating the parallax 
parameters into the analysis \citep{batista11,skowron11,han16}. We note that if one has 
an estimate for both $\pi_{\rm E}$ and $\theta_{\rm E}$, the physical lens parameters 
(lens mass $M_{\rm tot}$ and $D_{\rm L}$) can be determined by 
\begin{equation}
M_{\rm tot} = {\theta_{\rm E} \over \kappa\pi_{\rm E}};~~~~~D_{\rm L} = {{\rm AU} \over \pi_{\rm E}\theta_{\rm E} + \pi_{\rm S}},
\label{eq4}
\end{equation}
where $\kappa = 4G/(c^{2}{\rm AU}) \sim 8.14~{\rm mas}~M_{\odot}^{-1}$ and $\pi_{\rm S} = {\rm AU}/D_{\rm S}$.

The duration of the event is $t_{\rm d} = 2(1 - u_{0}^{2})^{1/2}t_{\rm E} \sim 85~{\rm days}$, 
which covers a significant portion of Earth's orbit period. Hence, we additionally 
fit the light curve by introducing the above higher-order parameters to the standard model. 
In this modeling, we also test $u_{0} > 0$ and $u_{0} < 0$ solutions to account for the 
``ecliptic degeneracy'' \citep{skowron11}. From the analysis, we find that the orbital parameters 
are poorly constrained. In particular, we find that some of the MCMC trials are in the regime of 
$\beta > 1$. Here $\beta$ is the ratio of transverse kinetic to potential energy given by 
\begin{equation}
\beta \equiv \left({{\rm KE} \over {\rm PE}}\right)_{\perp} = {{\kappa}M_{\odot}{\rm yr}^{2} \over 8\pi^{2}}{\pi_{\rm E} \over \theta_{\rm E}}{s^{3}{\gamma}^{2} \over (\pi_{\rm E} + \pi_{\rm S}/\theta_{\rm E})^{3}},
\label{eq5}
\end{equation}
where we adopt $\pi_{\rm S} = 0.128$ mas for the calculation based on the
clump distance in this direction \citep{nataf13}. We know a priori that this ratio 
must be less than unity, i.e., $\beta < 1$, to be a bounded system \citep{dong09}. Hence, 
we exclude the trials that show $\beta > 1$ as physically unrealistic solutions.

\begin{figure}[th]
\epsscale{0.9}
\plotone{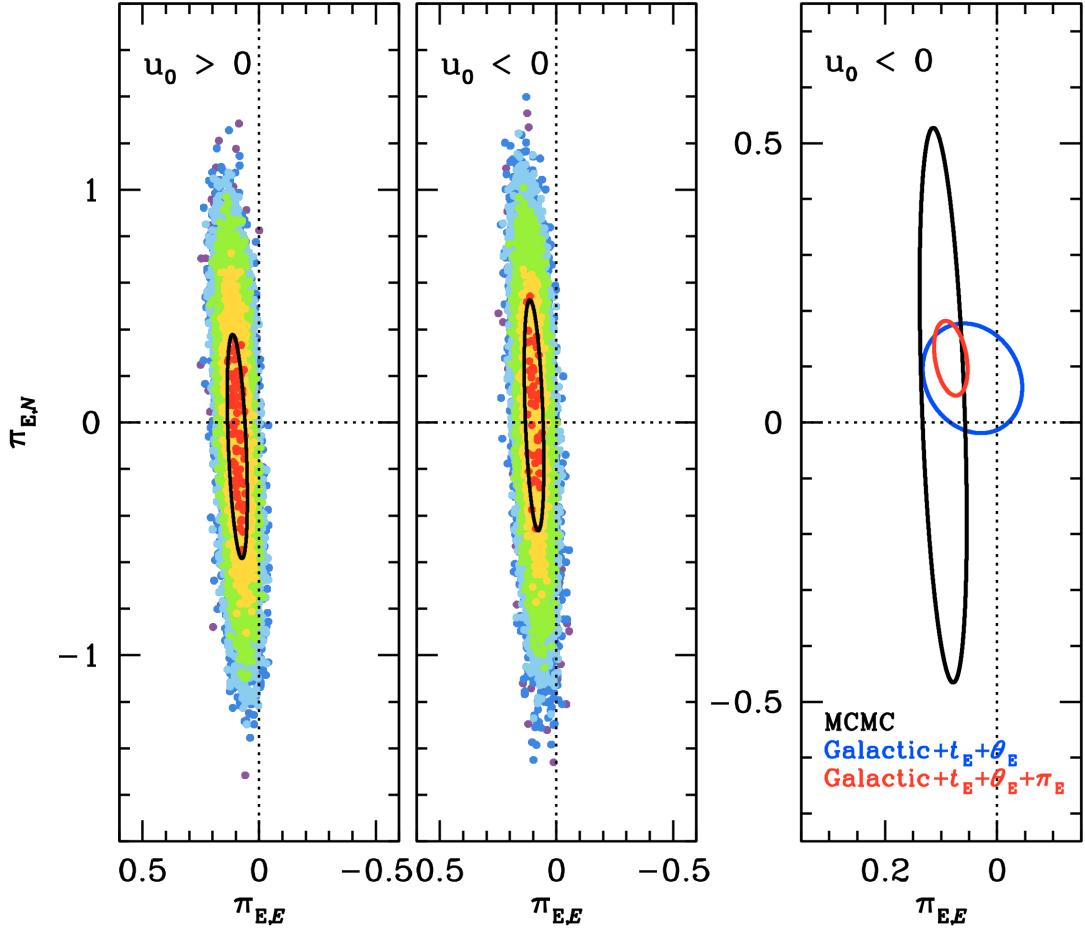}
\caption{\label{fig:four}
$\Delta\chi^{2}$ surfaces in $(\pi_{{\rm E}, N}, \pi_{{\rm E}, E})$ plane. 
The left and middle distributions are drawn from the two solutions ($u_{0} > 0$ and $u_{0} < 0$). 
Except that $n = 1$, the color coding is identical to that of Figure~\ref{fig:two}. 
In each panel, the black curve is the error contour $(\Delta\chi^2 = 1)$ extracted from the MCMC fit.
The right panel shows the evolution of the parallax vector ($u_0 < 0$) depending on the priors. 
The blue curve is the posterior distribution $(-2\Delta\ln{\cal L} = 1)$ derived from 
the timescale and angular Einstein radius constraints, while the red curve is 
based on the additional parallax constraint (see Section 4). 
}
\end{figure}

The results are presented in Table~\ref{table:one}. We find that the fit improvement from these higher-order effects 
is $\Delta\chi^2 \sim 9$, 
which has a Gaussian false probability (from four extra degrees of freedom) of just $(1 + \Delta\chi^2/2)\exp(-\Delta\chi^2/2) = 6\%$. 
This low level of improvement implies that it is difficult to directly determine the characteristics 
of the lens system from the microlensing fit parameters alone. Hence, we constrain the physical lens 
properties from a Bayesian analysis based on Galactic models. At this point, one might suggest that 
it is needless to introduce the measured higher-order effects to the analysis, but rather the lens system 
should be estimated only from the standard solution. However, this point of view is not always correct. 
As discussed in \citet{han16}, we know a priori that all binary microlenses have both finite microlens 
parallax and finite lens orbital motion. This indicates that even in such low- or non-measurement cases, 
well-constrained higher-order parameters can include considerable information about the lens system. 
Hence, they can play an important role in constraining the physical lens properties from statistical 
analyses. In our case, we find that the east component of parallax vector $\pi_{{\rm E},E}$ is well 
constrained for both the $u_{0} > 0$ and $u_{0} < 0$ solutions, although the error of the north component 
$\pi_{{\rm E},N}$ is considerable (see Figure~\ref{fig:four}). In addition, we find that for both 
parallax solutions (as well as the standard, no-parallax solution), the seven standard parameters 
(except the sign of $u_{0}$) are consistent within $1\sigma$, which implies that the existence of 
multiple solutions does not significantly affect the statistical expectation of the lens system. 
Therefore, in what follows, we show results of the Bayesian analysis only for $u_0 < 0$ solution. 
However, we note that the results for the $u_0>0$ solution are nearly identical.

\section{Physical Parameters}

\begin{figure}[th]
\epsscale{0.9}
\plotone{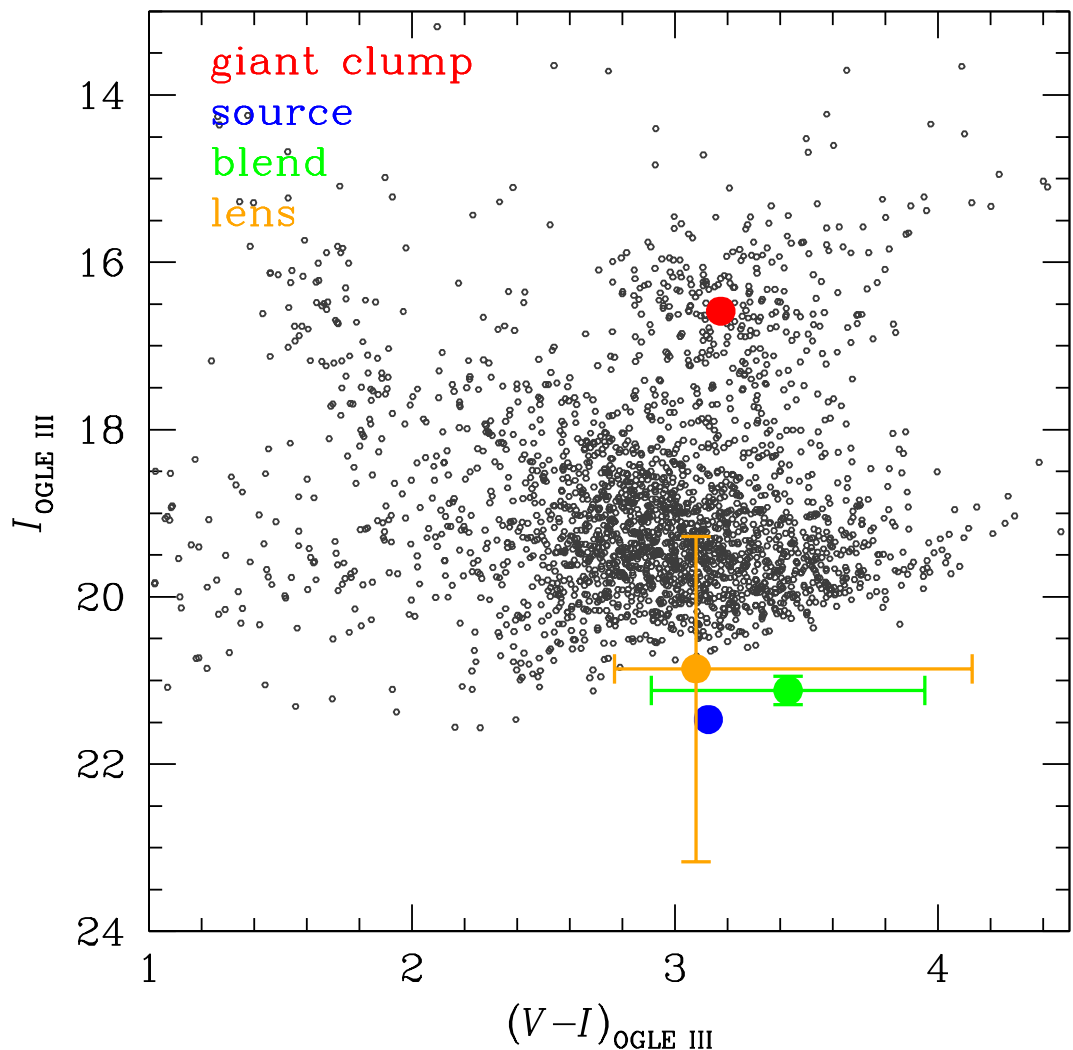}
\caption{\label{fig:five}
Calibrated color-magnitude diagram of field stars around KMT-2017-BLG-0165.
The locations of microlensed source and giant clump centroid (GC) are marked 
by the blue and red points, respectively. The orange point shows the color and magnitude 
expected on the basis of the mass and distance derived from the Bayesian analysis. 
The green point indicates the position of the blended light estimated from the CFHT $r$- and $i$-band images. 
The astrometric position of the blend is well aligned with the source. See Figure~\ref{fig:seven}. 
Hence it is plausible that the blended light is due to the lens.  See text.
}
\end{figure}

Our first step for constraining the physical lens parameters is to determine the angular Einstein 
radius $\theta_{\rm E}$. For this, we adopt the method of \citet{yoo04}. We derive the source 
color $(V-I)_{\rm S}$ and brightness $I_{\rm S}$ from the model using the KMTNet star catalog 
calibrated to the OGLE-III photometry map \citep{szymanski11}\footnote{We use the pyDIA reduction 
for constructing the KMTNet star catalog.}. We then measure the source offset from the giant clump 
centroid (GC) in the color-magnitude diagram (CMD): $\Delta(V-I,~I) = (V-I,~I)_{\rm S} - (V-I,~I)_{\rm GC} = 
(2.53\pm0.04,~18.43\pm0.02) - (2.58\pm0.03,~13.54\pm0.02) = (-0.05\pm0.05,~4.89\pm0.03)$. 
Figure~\ref{fig:five} shows the positions of source and GC in the CMD. 
We find the dereddened source position as 
\begin{equation}
(V-I,~I)_{0,{\rm S}} = \Delta(V-I,~I) + (V-I,~I)_{0,{\rm GC}} = (1.01\pm0.09,~19.25\pm0.09),
\label{eq6}
\end{equation}
where $(V-I,~I)_{0,{\rm GC}} = (1.06\pm0.07, 14.37\pm0.09)$ is the dereddened GC position 
adopted from \citet{bensby13} and \citet{nataf13}, respectively. We convert our 
estimated $(V-I)_{0,{\rm S}}$ to $(V-K)_{0,{\rm S}} = 2.30\pm0.09$ using 
the $VIK$ relation \citep{bessell88}, and then derive the angular source radius   
\begin{equation}
\theta_{*} = 0.63\pm0.06~\mu{\rm as}
\label{eq7}
\end{equation}
using the $(V-K)-\theta_{*}$ relation \citep{kervella04}. 
Here, the error in $\theta_{*}$ is estimated from 
the uncertainty of the source color measurement (4\%), centroiding the giant clump (7\%), 
and the color-surface brightness conversion (5\%). From the measured source radius $\rho_{*}$, 
we finally derive the angular Einstein radius 
\begin{equation}
\theta_{\rm E} = {\theta_{*} \over \rho_{*}} = 0.80\pm0.13~{\rm mas},
\label{eq8}
\end{equation}
which corresponds to the geocentric lens-source relative proper motion
\begin{equation}
\mu_{\rm geo} = {\theta_{\rm E} \over t_{\rm E}} = 6.93\pm1.15~{\rm mas~yr^{-1}}.
\label{eq9}
\end{equation}

With the measured $t_{\rm E}$, $\theta_{\rm E}$, and $\pi_{\rm E}$ constraints, 
we conduct a Bayesian analysis by adopting the procedure and Galactic model of \citet{jung18}. 
We first create a large sample of lensing events that are randomly drawn from the Galactic model. 
For each trial event, we then evaluate the likelihood of the microlensing parameters 
$(t_{\rm E}, \theta_{\rm E}, \pi_{{\rm E},N}, \pi_{{\rm E},E})_{k}$ that are explicitly predicted 
by each combination of simulated lens and source properties. If all four of these parameters were uncorrelated, 
the likelihood of the event given the model could be estimated by a product of four Gaussian distributions. 
However, while the correlations between $(t_{\rm E}, \theta_{\rm E})$, $(t_{\rm E}, \pivec_{\rm E})$, 
and $(\theta_{\rm E}, \pivec_{\rm E})$ are all quite weak, the correlation between 
$(\pi_{{\rm E},N}, \pi_{{\rm E},E})$ can be quite strong \citep{gould04}. Therefore, we estimate the likelihood 
as the product of a bivariate Gaussian of $(\pi_{{\rm E},N}, \pi_{{\rm E},E})$ 
with two univariate Gaussians of $t_{\rm E}$ and $\theta_{\rm E}$. For this, we evaluate 
the $\chi^2$ difference between the simulated and the measured values as 
\begin{equation}
\chi_{{\rm gal}, k}^{2} = \chi_{k}^{2}(t_{\rm E}) + \chi_{k}^{2}(\theta_{\rm E}) + \chi_{p, k}^{2};~~~~~
\chi_{p, k}^{2} = \sum(a_k - a_0)_{i}c_{ij}^{-1}(a_k - a_0)_{j}, 
\label{eq10}
\end{equation}
where $\mathbf{a}_{k} = \pivec_{{\rm E}, k} =  (\pi_{{\rm E},N},~\pi_{{\rm E},E})_{k}$, and $a_{0}$ 
and $c$ are the mean value of $\pivec_{\rm E}$ and its covariance matrix extracted from the MCMC, 
respectively. We then estimate the relative likelihood of the event by 
\begin{equation}
{\rm P}_{k} = {\rm exp}(-\chi_{{\rm gal}, k}^{2}/2)\times\Gamma_{k},
\label{eq11}
\end{equation}
where $\Gamma_{k} \propto \theta_{{\rm E},k}\mu_{k}$ is the microlensing event rate. Finally, 
we explore the likelihood distributions of the lens properties 
from all trial events using ${\rm P}_{k}$ as a prior. We note that to show the contributions 
of individual constraints for estimating the physical lens properties, we additionally explore 
the distributions (1) with only the $t_{\rm E}$ constraint and (2) with the $t_{\rm E}$ and 
$\theta_{\rm E}$ constraints.

\begin{figure}[th]
\epsscale{0.9}
\plotone{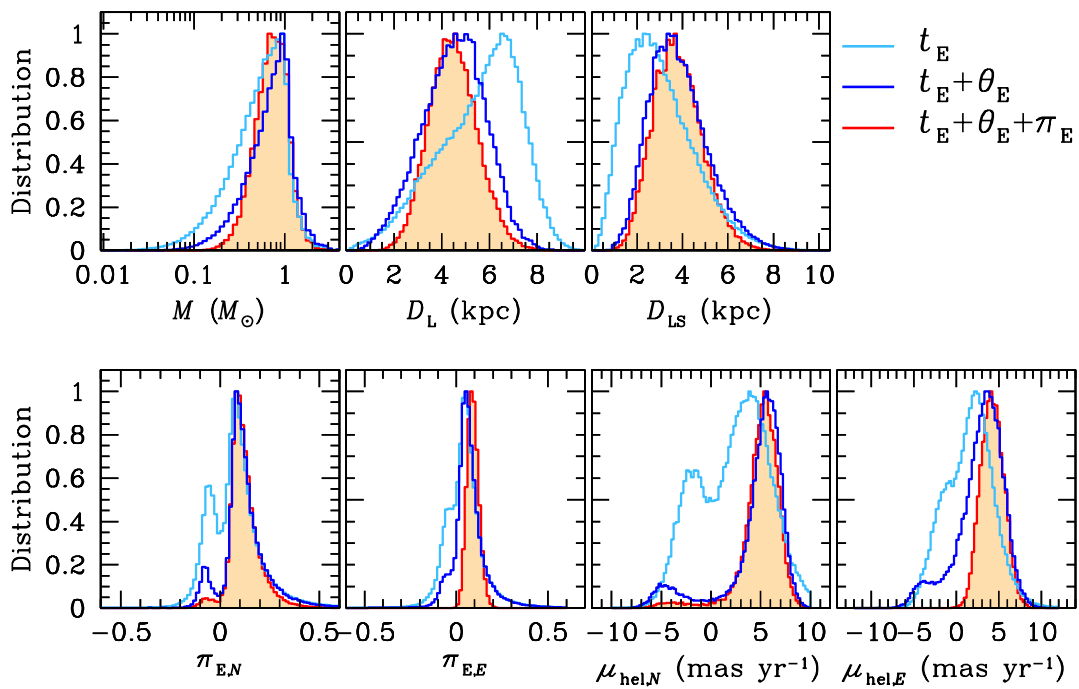}
\caption{\label{fig:six}
Likelihood distributions of the lens properties derived from the Bayesian analysis. 
In each panel, the three curves with light blue, blue, and red colors show the distributions derived 
from the timescale constraint, the timescale and angular Einstein constraints, 
and the additional parallax constraint, respectively. The distributions are for the $u_{0} < 0 $ solution, 
but those for the $u_{0} > 0$ and no-parallax solutions are virtually identical.       
}
\end{figure}

The results are shown in Figure~\ref{fig:six}. We find that the measured 
$\pi_{\rm E}$ and $\theta_{\rm E}$ provide a strong constraint on the distributions 
(see the right panel in Figure~\ref{fig:four}). The median values of the lens host-mass 
$M_{1}$ and distance $D_{\rm L}$ with $68\%$ confidence intervals are 
\begin{equation}
M_{1} = {0.76}_{-0.27}^{+0.34}~M_{\odot},~~~~~D_{\rm L} = {4.53}_{-0.98}^{+1.04}~{\rm kpc},  
\label{eq12}
\end{equation}
respectively. The corresponding heliocentric source proper motion relative to the lens are
\begin{equation}
\mu_{\rm hel} = \left|\muvec_{\rm geo} + {\mathbf v}_{\oplus,\perp}{\pi_{\rm rel} \over {\rm AU}}\right| = 6.93_{-1.54}^{+1.38}~{\rm mas\,yr^{-1}},  
\label{eq13}
\end{equation}
where ${\mathbf v}_{\oplus,\perp} = (v_{\oplus, N}, v_{\oplus, E}) = (2.21, 10.42)~{\rm km\,s^{-1}}$ 
is Earth's projected velocity at $t_{0}$. Combined with the lens-source distance 
$D_{\rm LS} = 3.80_{-1.00}^{+1.13}~{\rm kpc}$, these indicate that the lens host 
is a Sun-like star located in the Galactic disk. The mass of the planet and its 
projected separation from the host are then estimated by 
\begin{equation}
M_{2} = qM_{1} = 34_{-12}^{+15}~M_{\oplus},~~~~~a_{\bot} = s{D_{\rm L}}{\theta_{\rm E}} = 3.45_{-0.95}^{+0.98}~{\rm AU}. 
\label{eq14}
\end{equation}
That is, the planet is a cold super-Neptune lying outside the snowline of the host, 
i.e., $a_{sl} = 2.7\,{\rm AU}(M/M_{\odot}) \sim 2.1\,{\rm AU}$. We summarize the estimated 
lens properties including the measured $\theta_{\rm E}$, $\mu_{\rm geo}$, and $\muvec_{\rm hel}$ in Table~\ref{table:two}.

We now investigate whether these Bayesian estimates of the lens physical properties 
imply a lens flux that is consistent with the blended light at the position of the 
microlensed source. This comparison requires two distinct steps. First, we must place
the lens-flux estimates on the calibrated CMD (Figure~\ref{fig:five}).
Second, we must make a refined estimate of the blended light and
place this estimate on the same calibrated CMD. Based on the lens host-mass, 
we first estimate the absolute brightness of the lens in $I$ band as 
$M_{I} = 5.37_{-1.58}^{+2.31}$ \citep{pecaut13}. Because the lens distance 
is $D_{\rm L} \sim 4.5~{\rm kpc}$, it is expected that the lens is probably behind most of the 
dust in the disk. Hence, we assume that the lens experiences similar reddening and extinction 
to those of the microlensed source. The lens position in the CMD is then estimated by   
\begin{equation}
(V-I, I)_{\rm L} = (V-I, I)_{0,\rm L}+(V-I, I)_{\rm GC}-(V-I, I)_{0,{\rm GC}} = (3.08_{-0.31}^{+1.05}, 20.86_{-1.58}^{+2.31}),  
\label{eq15}
\end{equation}
where $I_{0,{\rm L}} = M_{I} + 5{\rm log}(D_{\rm L}/{\rm pc}) - 5$ and 
$(V-I)_{0,\rm L} = 0.97_{-0.31}^{+1.05}$ is the intrinsic color of the lens adopted 
from \citet{pecaut13}.

The blended light cannot be accurately estimated either from the KMTNet reference image 
or from the source characteristics listed in the OGLE-III catalog \citep{szymanski11} 
that is used to identify microlensed sources by the KMTNet event-finder \citep{kim18}, 
because the source star is not resolved in either image. That is, the OGLE-III-based 
catalog star lies $0.6^{\prime\prime}$ from the microlensed source and thus 
this catalog star must be a blend of the source, the lens, as well as one or more stars that 
are substantially displaced from these objects. Similarly, there is no 
distinct ``star'' at the location of the event in the KMTNet reference image. 
However, we find that high-resolution images of KMT-2017-BLG-0165 were observed by the 
2016 CFHT-K2C9 Multi-color Microlensing Survey \citep{zang18}, a special survey designed 
to measure the colors of microlensed sources for K2's Campaign 9 microlensing survey 
\citep{henderson16} with the $g$-, $r$-, and $i$-band filters of the Canada-France-Hawaii 
Telescope (CFHT) on Mauna Kea. Therefore, we make use of these $0.187^{\prime\prime}$
pixel images, which have FWHM $\sim 0.5^{\prime\prime}$.

\begin{figure}[th]
\epsscale{0.9}
\plotone{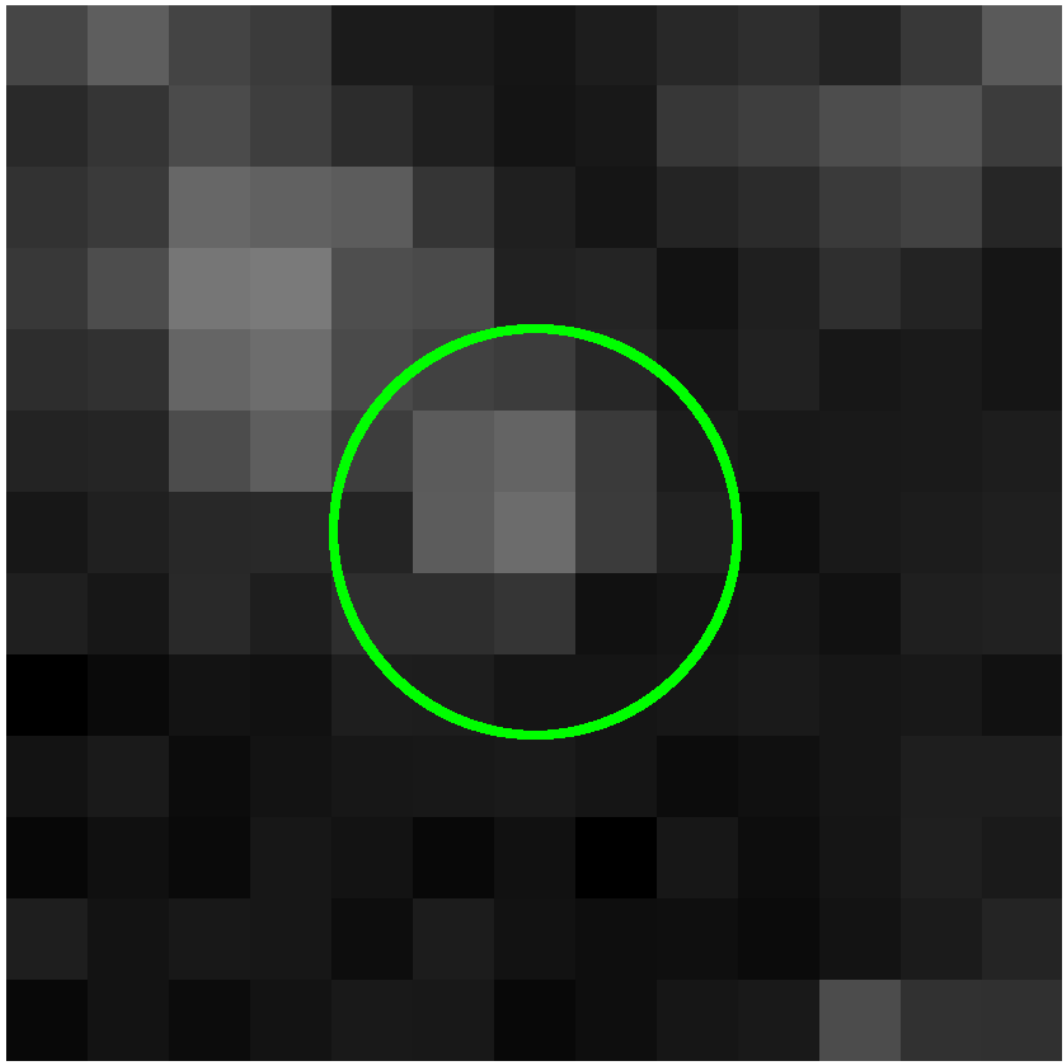}
\caption{\label{fig:seven}
$i$-band CFHT image within $2.4^{\prime\prime}\times2.4^{\prime\prime}$ around the event. 
The green circle indicates the source position derived from the astrometric transformation 
of a highly magnified KMTNet image.
}
\end{figure}

To estimate the blended light, we first identify the source position in the CFHT images from 
an astrometric transformation of a highly magnified KMTNet image. Figure~\ref{fig:seven} shows 
one of these images together with the position of the source circled in green. We find that 
the ``baseline object'' associated with the source and lens (and possibly other stars) is 
clearly resolved in the $r$- and $i$-band images. We then perform aperture photometry and 
align it to the calibrated OGLE-III system shown in Figure~\ref{fig:five}. From this, 
we find\footnote{The errors in the instrumental magnitudes are $\pm 0.15\,$mag and $\pm 0.08\,$mag 
for CFHT $r$- and $i$-band, respectively. However, because the transformations (in the neighborhood 
of the observed instrumental color) from instrumental $(r,i)$ to standard $(V,I)$ are 
$V = 1.60 r - 0.60 i + {\rm const}$ and $I = 1.16 i - 0.16 r + {\rm const}$, the difference 
between the errors in $V$ and $I$ are larger than between $r$ and $i$.} 
$V_{\rm base} = 23.82\pm0.24$ and $I_{\rm base} = 20.53\pm0.09$. Subtracting the source flux 
(from the model fit) yields the $(V-I)$ color and $I$-band magnitude of the blended light as 
$(V-I, I)_{\rm b}=(3.43\pm0.52, 21.12\pm0.17)$. We then plot this value in green on the CMD 
(see Figure~\ref{fig:five}). We see from Figure~\ref{fig:five} that the blended flux (green) 
is quite consistent (within $1\sigma$) with the flux predicted for the lens (orange). 
The relation of the blended light to the lens could be further investigated 
using adaptive optics (AO) follow up.

\section{Discussion}

KMT-2017-BLG-0165Lb is a planet with a mass-ratio of $q = 1.3\times10^{-4}$, 
which is near the \citet{suzuki16} break, $q_{\rm br} = 1.7\times10^{-4}$. 
In this context, we investigate the location ($q_{\rm br}$) and strength ($\zeta$) 
of the mass-ratio break from the distribution of microlensing planets in the region of the 
break: $q\leq 3.0\times 10^{-4}$. For this, we first review the literature and find the 
planets whose mass ratios are securely measured (without any strong degeneracy) and 
lie below $q = 3.0\times10^{-4}$. We find 15 planets (including KMT-2017-BLG-0165Lb) 
that satisfy these criteria\footnote{The low mass-ratio $(q < 10^{-4})$ portion of this 
sample is identical to the sample in \citet{udalski18}.}. Table~\ref{table:three} 
gives the main characteristics of these planets. We note that there is another 
planetary event, OGLE-2017-BLG-0173 \citep{hwang18}, whose mass ratio falls in the defined 
range. However, the event suffers from large uncertainties in the mass ratio measurement 
due to a discrete degeneracy between two classes of solutions, i.e., $q = (2.5, 6.5)\times10^{-5}$.
Whenever two solutions with different mass ratios are roughly equally 
consistent with the data, including such planets in the analysis would mean that their 
role depends on the priors. In our case, however, we do not have independent (prior) 
knowledge of the mass-ratio function, and this is what we are trying to measure. 
Therefore, we exclude the event in order to obtain reliable independent results.

\begin{figure}[th]
\epsscale{0.9}
\plotone{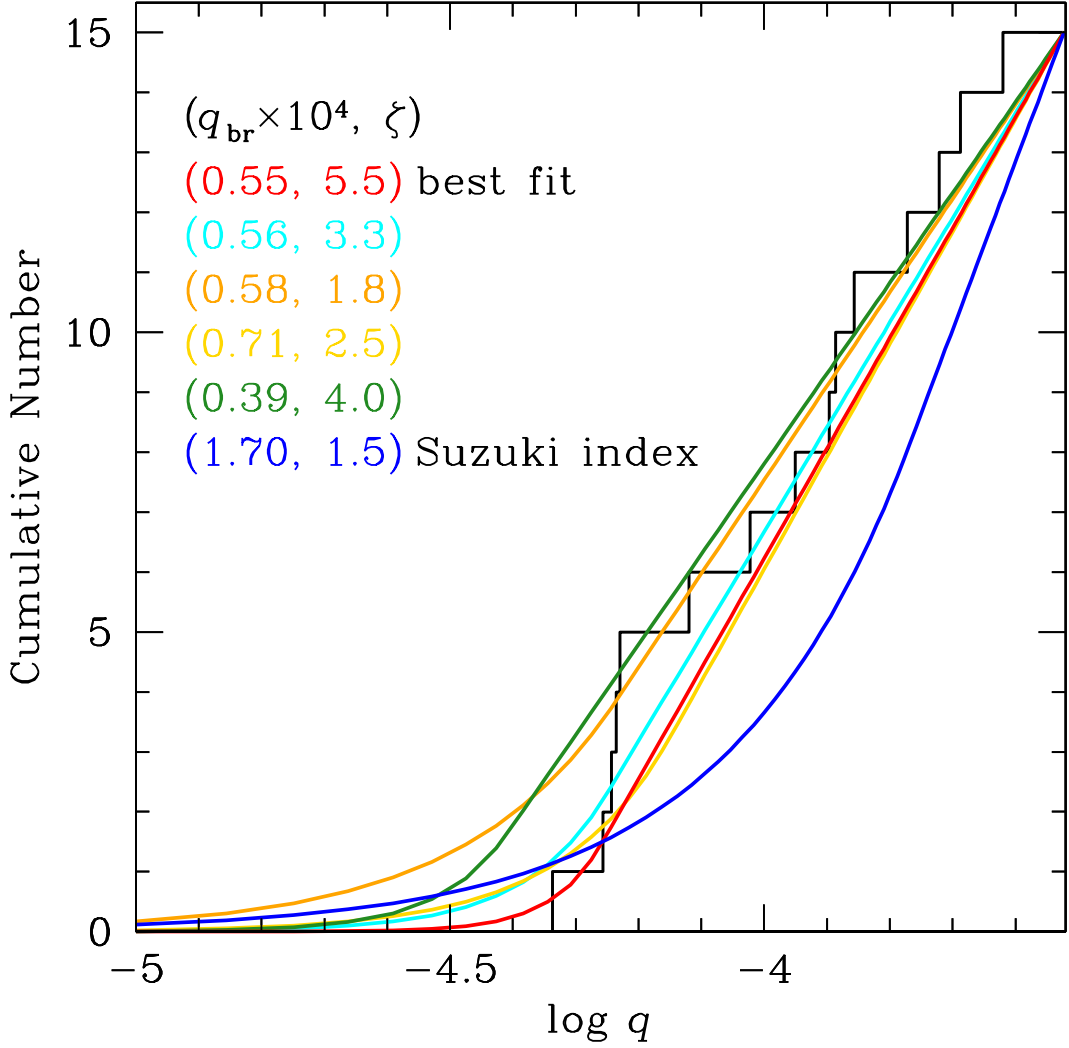}
\caption{\label{fig:eight}
Cumulative distribution of the mass ratio $q$ in the regime of $q < 3\times10^{-4}$. 
The black curve is the actual cumulative distribution of fifteen microlensing planets. 
The blue curve is the predicted distribution based on the result of \citet{suzuki16}, 
while the remaining curves are some representative (including the best fit) distributions derived from the likelihood 
${\cal L}$ analysis. 
}
\end{figure}

We find that the cumulative distribution in our sample domain 
$(4\times 10^{-5}<q<3\times 10^{-4})$ is primarily characterized by a straight line, 
which corresponds to a uniform distribution in ${\rm log}~q$ over $\Delta\log q=0.875$ decades, 
i.e., $d N_{\rm obs}/d\log q = {\rm const}$ (see Figure~\ref{fig:eight}). 
The most notable feature within this overall trend 
is a ``pile up'' of four planets in the short interval ($\Delta\log q=0.030$ decades) 
at $5.5\times 10^{-5}<q<5.9\times 10^{-5}~(-4.26 < {\rm log}~q < -4.23)$.

Next, we calculate the cumulative distribution for a broken power law.  
We begin by considering that the intrinsic occurrence rate of planets 
across the whole range that is being probed follows a power-law function 
$f_{0}({\rm log}~q) = dN_{0}/d{\rm log}~q \sim q^{\alpha}$. 
We must also consider the sensitivity of microlensing experiments to planets, 
which we assume is a power law\footnote{This assumption is well-motivated 
by various planet sensitivity studies 
\citep{gould10,cassan12,suzuki16,zhu17}, i.e., $S(q) \sim q^{\beta}$.}. 
The observed planet frequency is then
\begin{equation}
f_{\rm obs}({\rm log}~q) = {dN_{\rm obs} \over d{\rm log}~{q}} = f_{0}({\rm log}~q)S(q) \sim q^{\alpha+\beta}.
\label{eq16}
\end{equation}
For our calculation, we adopt that the cumulative distribution is linear, i.e., 
constant number of detections in each bin of equal ${\rm log}~q$ \citep{mroz17a}. 
This means that the frequency is flat, i.e., $f_{\rm obs}({\rm log}~q) \sim q^{0}$. 
That is, $\alpha = -\beta$. We then apply the broken power-law function as a form of  
\begin{equation}
g_{0}({\rm log}~q) = K\left({q \over q_{\rm br}}\right)^{\alpha}~(q > q_{\rm br});~~~~~ 
g_{0}({\rm log}~q) = K\left({q \over q_{\rm br}}\right)^{\alpha+\zeta}~(q < q_{\rm br}), 
\label{eq17}
\end{equation}
where $K$ is a normalization constant. 
The observed planet frequency for the broken power-law 
$g_{\rm obs}({\rm log}~q) = g_{0}({\rm log}~q)S(q)$ is then given by 
\begin{equation}
g_{\rm obs}({\rm log}~q) = K~(q > q_{\rm br});~~~~~ 
g_{\rm obs}({\rm log}~q) = K\left({q \over q_{\rm br}}\right)^{\zeta}~(q < q_{\rm br}). 
\label{eq18}
\end{equation}
Finally, we derive the cumulative distribution function 
$G_{\rm obs}({\rm log}~q) = \int_{0}^{q}g_{\rm obs}({\rm log}~q')d{\rm log}~q'$ as 
\begin{equation}
G_{\rm obs}({\rm log}~q) = K\left[{1 \over \zeta} + {\rm ln}\left({q \over q_{\rm br}}\right)\right]~(q > q_{\rm br});~~~~~ 
G_{\rm obs}({\rm log}~q) = {K \over \zeta}\left({q \over q_{\rm br}}\right)^{\zeta}~(q < q_{\rm br}). 
\label{eq19}
\end{equation}
Here, we find the constant $K$ by matching this function to the actual cumulative 
distribution, i.e.,  
\begin{equation}
K = {G_{\rm max} \over 1/\zeta + {\rm ln}(q_{\rm max}/q_{\rm br})},
\label{eq20}
\end{equation}
where $G_{\rm max} = G_{\rm obs}({\rm log}~q_{\rm max}) = G_{\rm obs}[{\rm log}~(3\times10^{-4})] = 15$. 
Therefore, for a given mass ratio $q$, the predicted cumulative distribution $G_{\rm obs}({\rm log}~q)$ 
can be determined as a function of $\zeta$ and $q_{\rm br}$.

\begin{figure}[th]
\epsscale{0.9}
\plotone{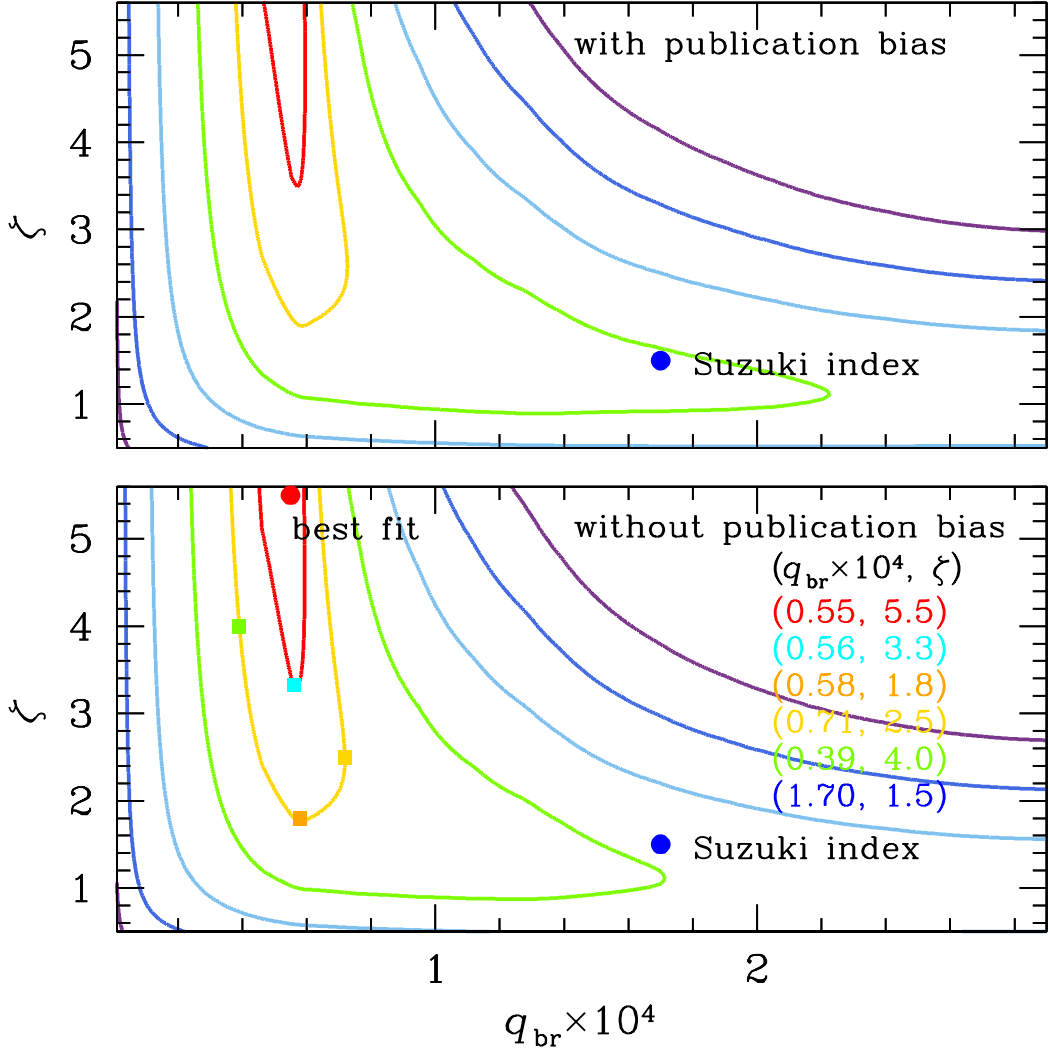}
\caption{\label{fig:nine}
$-2\Delta\ln{\cal L}$ contour in $(q_{\rm br},~\zeta)$ plane. 
The lower and upper panels show the distributions derived without and with the ``publication bias'' adjustment, 
respectively. The red point is the location of best-fit model, while the blue point is the 
location of \citet{suzuki16} index. The squares with different colors are the positions of 
some representative models shown in Figure~\ref{fig:eight}. 
The color coding is identical to that of Figure~\ref{fig:two}, except that $n = 1$. 
}
\end{figure}

Next, we calculate the likelihood ${\cal L}$ of the observed 15 planets given 
model functions defined by two parameters $(q_{\rm br},\zeta)$. In Figure~\ref{fig:nine}, 
we show the contours of the result according to $-2\Delta\ln{\cal L} = (1,4,9,16,25,36)$. 
The cumulative distributions for some representative models are also shown in Figure~\ref{fig:eight}.
The best fit $(q_{\rm br}\times 10^4,\zeta)=(0.55,5.5)$ is shown as a red curve. 
Because $q_{\rm br}=0.55\times 10^{-4}$ is below all but two of the observed planets, 
this model matches the observed (flat) distribution quite well. We find that the model 
accounts for the absence of observed planets $q<4.6\times 10^{-4}$ by an extremely 
strong power-law break, $\zeta=5.5$. As shown in Figure~\ref{fig:nine}, we also find 
that the probable (i.e., ``$1\,\sigma$'') models all have this same break point 
$q_{\rm br}\simeq 0.55\times 10^{-4}$. The lowest strength for probable models is $\zeta=3.3$ 
(see the cyan curve in Figure~\ref{fig:eight}). That is, all of these models require 
an extremely sharp power-law break, which corresponds approximately to a cut off 
just below $q_{\rm br}$.

We note that from our sample planets, we identify a bias, so-called ``publication date'' 
bias \citep{mroz17a}. We find that on average, the three planets with $q > 1.7\times10^{-4}$ 
show a five year delay for their publications relative to their discovery years, 
while the planets at lower $q$ show only a one-year delay. In order to check 
whether our result can be affected by this bias, we additionally calculate 
the likelihood distribution with the ``publication bias'' adjustment, 
for which we weight the counts of those three planets by a factor of 1.5
\footnote{Or, equivalently, in our likelihood fits, we assume a reduced 
probability of discovery by a factor 2/3.}, i.e., $G_{\rm max} = 16.5$. 
From this, we find that the derived distribution is virtually identical to that derived 
without the adjustment (see Figure~\ref{fig:nine}).

Our result is quite different from that of \citet{suzuki16}. First, the break is at much 
lower mass ratio than their best value: $q_{\rm br}=1.7\times10^{-4}$. Second, the break 
is also much stronger than their adopted range $\zeta=1.5^{+0.5}_{-0.4}$. We find that both 
of these characteristics follow directly from the observed cumulative distribution shown in 
Figure~\ref{fig:eight}. As noted above, the straight-line cumulative distribution corresponds 
to detections that are uniform in $\log q$, and these detections simply stop below the 
$x$-intercept of this straight line. This implies more of a ``cut off'' than a break in the 
power-law index. Within the context of broken-power-law models, this ``cut off'' is then 
mathematically manifested as a very steep break in the index just above the lowest-$q$ detection.

The slope we find is also different from the estimate by \citet{udalski18}, who derived 
a value similar to that of \citet{suzuki16}. This difference is due to the difference in the 
configuration used for the analysis. Instead of looking for the position of the break, 
\citet{udalski18} assumed that the break is placed above $q > 10^{-4}$ and that the mass-ratio function 
below $q < 10^{-4}$ has the form of a power law. With these assumptions, they only investigated 
the index of this power law using the seven planets in the $q < 10^{-4}$ regime. 
By contrast, we seek to not only identify the position of the break in the power law 
but also find the change of the power-law slope between mass ratios above and below the break.
As presented above, our estimated break is at $q_{\rm br}=0.55\times 10^{-4}$, corresponding to 
the middle of the \citet{udalski18} sample. This contradicts their assumption 
that the break occurs above $q > 10^{-4}$. Therefore, it is inevitable 
that the conclusions are different.

In principle, the complete absence of detections for $q<0.4\times 10^{-4}$ could be 
due to a catastrophic decline in microlensing sensitivity to such low-mass-ratio planets. 
However, \citet{udalski18} studied the sensitivity of the seven events with $q<10^{-4}$, 
and found that microlensing studies can probe planets in the regime quite well, where 
(as shown in Figure~\ref{fig:eight}) there are no actual detected planets. Hence, this 
suggests that the absence of detected planets in this mass-ratio regime reflects 
their paucity in nature.

We next ask: is the ``pile up''of four planets within $\Delta\log q = 0.03$, mentioned 
above, real? It is difficult to devise reliable statistical tests for such posteriori ``features''. 
A simple Kolmogorov-Smirnov (KS) test ($D=0.2$ for an $n=15$ sample) yields a false probability 
of $p=12\%$. This is certainly not strong enough to reject the class of broken-power-law models. 
On the other hand, there is no a priori reason that the planet mass-ratio frequency should
follow a broken power law. In particular, physically based models could plausibly account 
for a pile-up at Neptune-like mass ratios because this is near to the point that a rock/ice 
core can start to accumulate a gaseous envelop. Therefore, we suggest that the apparent 
``pile up'' in Figure~\ref{fig:eight} may be real and warrants further investigation as more 
statistics are accumulated.

\acknowledgments
This research has made use of the KMTNet system operated by the Korea 
Astronomy and Space Science Institute (KASI) and the data were obtained at 
three host sites of CTIO in Chile, SAAO in South Africa, and SSO in Australia. 
This research uses data obtained through the Telescope Access Program (TAP), 
which has been funded by the National Astronomical Observatories of China, 
the Chinese Academy of Sciences (the Strategic Priority Research Program 
``The Emergence of Cosmological Structures'' Grant No. XDB09000000), and 
the Special Fund for Astronomy from the Ministry of Finance. This work was 
partly supported by the National Science Foundation of China (Grant No. 11333003, 
11390372 and 11761131004 to SM). This work was performed in part under contract with 
the California Institute of Technology (Caltech)/Jet Propulsion Laboratory (JPL) 
funded by NASA through the Sagan Fellowship Program executed by the NASA Exoplanet 
Science Institute. C. Han was supported by grant 2017R1A4A1015178 of the National Research 
Foundation of Korea. Work by AG were supported by AST-1516842 from 
the US NSF. AG were supported by JPL grant 1500811. 
AG is supported from KASI grant 2016-1-832-01. AG received support from the
European  Research  Council  under  the  European  Union's
Seventh Framework Programme (FP 7) ERC Grant Agreement n. [321035]. 
Work by MTP was partially supported by NASA grants NNX16AC62G and NNG16PJ32C.

\begin{deluxetable}{lrrrrr}
\tablecaption{Lensing Parameters\label{table:one}}
\tablewidth{0pt}
\tablehead{
\multicolumn{1}{l}{Parameters} &
\multicolumn{1}{c}{Standard} &
\multicolumn{2}{c}{Orbit+Parallax} \\
\multicolumn{2}{c}{} & 
\multicolumn{1}{c}{$u_0 > 0$} & 
\multicolumn{1}{c}{$u_0 < 0$} 
}
\startdata
$\chi^2$/dof               & 30210.9/30182        &    30202.1/30178      &   30201.6/30178        \\
$t_0$ (${\rm HJD'}$)       & 7853.73$\pm$ 0.063   & 7853.72$\pm$ 0.065    & 7853.71 $\pm$ 0.067   \\
$u_0$                      &  0.034 $\pm$ 0.010   &  0.036 $\pm$ 0.011    &  -0.035 $\pm$ 0.011    \\
$t_{\rm E}$ (days)         & 43.105 $\pm$ 0.771   & 41.775 $\pm$ 0.846    &  42.129 $\pm$ 0.932    \\
$s$                        &  0.951 $\pm$ 0.009   &  0.953 $\pm$ 0.012    &   0.953 $\pm$ 0.013    \\
$q$ ($10^{-4}$)            &  1.392 $\pm$ 0.036   &  1.332 $\pm$ 0.086    &   1.348 $\pm$ 0.090    \\
$\alpha$ (rad)             &  5.821 $\pm$ 0.029   &  5.817 $\pm$ 0.032    &  -5.821 $\pm$ 0.033    \\
$\rho_\ast$ ($10^{-4}$)    &  7.863 $\pm$ 0.759   &  7.709 $\pm$ 1.029    &   7.834 $\pm$ 1.072    \\
$\pi_{{\rm E}, N}$         & --                   &  0.049 $\pm$ 0.457    &  -0.099 $\pm$ 0.471     \\
$\pi_{{\rm E}, E}$         & --                   &  0.088 $\pm$ 0.041    &   0.108 $\pm$ 0.040     \\
$ds/dt$ (yr$^{-1}$)        & --                   &  0.416 $\pm$ 1.183    &  -0.745 $\pm$ 1.204     \\
$d\alpha/dt$ (yr$^{-1}$)   & --                   &  0.369 $\pm$ 0.272    &   0.531 $\pm$ 0.276    \\    
$f_{\rm s}$                &  0.038 $\pm$ 0.003   &  0.040 $\pm$ 0.003    &   0.040 $\pm$ 0.003    \\  
$f_{\rm b}$                &  0.125 $\pm$ 0.009   &  0.124 $\pm$ 0.009    &   0.124 $\pm$ 0.009      
\enddata 
\vspace{0.05cm}
\tablecomments{
${\rm HJD}'= {\rm HJD}-2450000$
}
\end{deluxetable}

\begin{deluxetable}{lr}
\tablecaption{Lens properties\label{table:two}}
\tablewidth{0pt}
\tablehead{
\multicolumn{1}{l}{Parameters} &
\multicolumn{1}{c}{Values} 
}
\startdata
$\theta_{\rm E}$ (mas)                               &   0.80$\pm$0.13             \\ 
$\mu_{\rm geo}$ $({\rm mas}\,{\rm yr}^{-1})$         &   6.93$\pm$1.15             \\ 
$\mu_{\rm hel}$ $({\rm mas}\,{\rm yr}^{-1})$         &   $6.93_{-1.54}^{+1.38}$    \\ 
$\mu_{{\rm hel},N}$ $({\rm mas}\,{\rm yr}^{-1})$     &   $5.41_{-1.65}^{+1.35}$    \\ 
$\mu_{{\rm hel},E}$ $({\rm mas}\,{\rm yr}^{-1})$     &   $4.33_{-1.36}^{+1.42}$    \\ 
$M_{1}$ $(M_{\odot})$                                &   $0.76_{-0.27}^{+0.34}$    \\ 
$M_{2}$ $(M_{\oplus})$                               &   $34_{-12}^{+15}$          \\ 
$D_{\rm L}$ (kpc)                                    &   $4.53_{-0.98}^{+1.04}$   \\
$a_{\bot}$ (AU)                                      &   $3.45_{-0.95}^{+0.98}$   
\enddata 
\vspace{0.05cm}
\end{deluxetable}

\begin{landscape}
\begin{deluxetable}{lrrrrrrr}
\tablecaption{Characterisitcs of planets\label{table:three}}
\tablewidth{0pt}
\tablehead{
\multicolumn{1}{l}{Event} &
\multicolumn{1}{c}{$q (10^{-4})$} & 
\multicolumn{1}{c}{$s$} & 
\multicolumn{1}{c}{$M_{\rm p}/M_{\oplus}$} & 
\multicolumn{1}{c}{$M_{\rm h}/M_{\odot}$}  & 
\multicolumn{1}{c}{$D_{\rm L}/{\rm kpc}$}  & 
\multicolumn{1}{c}{$a_{\perp}/{\rm AU}$}   &   
\multicolumn{1}{c}{Discovery paper}   
}
\startdata
OGLE-2013-BLG-0341   &   0.46 &  0.81  &   2.00   &  0.15  & 1.16  &   0.88  & \citet{gould14}              \\
OGLE-2016-BLG-1195   &   0.55 &  0.98  &   1.43   &  0.08  & 3.91  &   1.16  & \citet{shvartzvald17}        \\
OGLE-2017-BLG-1434   &   0.57 &  0.98  &   4.48   &  0.23  & 0.87  &   1.18  & \citet{udalski18}            \\
MOA-2009-BLG-266     &   0.58 &  0.91  &  10.40   &  0.56  & 3.04  &   3.20  & \citet{muraki11}             \\
OGLE-2005-BLG-169    &   0.59 &  1.02  &  14.10   &  0.69  & 4.10  &   3.50  & \citet{gould06}$^{\dagger}$  \\
OGLE-2005-BLG-390    &   0.76 &  1.61  &   5.50   &  0.22  & 6.60  &   2.60  & \citet{beaulieu06}           \\
OGLE-2007-BLG-368    &   0.95 &  0.93  &  20.00   &  0.64  & 5.90  &   2.80  & \citet{sumi10}               \\
MOA-2007-BLG-192     &   1.20 &  1.12  &   3.30   &  0.06  & 1.00  &   0.62  & \citet{bennett08}            \\
MOA-2011-BLG-028     &   1.27 &  1.69  &  30.00   &  0.75  & 7.38  &   4.14  & \citet{skowron16}            \\    
OGLE-2012-BLG-0026   &   1.30 &  1.03  &  46.07   &  1.06  & 4.02  &   4.00  & \citet{han13}                \\
KMT-2017-BLG-0165    &   1.35 &  0.95  &  34.00   &  0.76  & 4.53  &   3.45  & this work                    \\
OGLE-2015-BLG-0966   &   1.69 &  1.12  &  21.00   &  0.38  & 2.50  &   2.10  & \citet{street16}             \\
OGLE-2012-BLG-0950   &   1.90 &  1.00  &  35.00   &  0.56  & 3.00  &   2.70  & \citet{koshimoto17}          \\
MOA-2012-BLG-505     &   2.05 &  1.13  &   6.70   &  0.10  & 7.21  &   0.91  & \citet{nagakane17}           \\
OGLE-2008-BLG-092    &   2.41 &  5.26  &  43.60   &  0.71  & 8.10  &  18.00  & \citet{poleski14}
\enddata 
\vspace{0.05cm}
\tablecomments{
${^\dagger}$ For OGLE-2005-BLG-169, the values are obtained from \citet{bennett15}, 
who refined the solution from follow-up observations of the lens and the source stars using the \textit{Hubble Space Telescope} (HST) Wide Field Camera 3 (WFC3).  
}
\end{deluxetable}
\end{landscape}

\begin{thebibliography}{99}

\bibitem[Alard \& Lupton(1998)]{alard98}
Alard, C., \& Lupton, Robert H. 1998, \apj, 503, 325

\bibitem[Albrow et al.(2009)]{albrow09}
Albrow, M. D., Horne, K., Bramich, D. M., et al. 2009, \mnras, 397, 2099


\bibitem[Batista et al.(2011)]{batista11}
Batista, V., Gould, A., Dieters, S., et al. 2011, \aap, 529, 102

\bibitem[Beaulieu et al.(2006)]{beaulieu06}
Beaulieu, J.-P., Bennett, D. P., \& Fouqu{\'e}, P. 2006, \nat, 439, 437

\bibitem[Bennett et al.(2015)]{bennett15}
Bennett, D. P., Bhattacharya, A., Anderson, J., et al. 2015, \apj, 808, 169

\bibitem[Bennett et al.(2008)]{bennett08}
Bennett, D. P., Bond, I. A., Udalski, A., et al. 2008, \apj, 684, 663

\bibitem[Bensby et al.(2013)]{bensby13}
Bensby, T., Yee, J. C., Feltzing, S., et al. 2013, \aap, 549, 147

\bibitem[Bessell \& Brett(1988)]{bessell88}
Bessell, M. S., \& Brett, J. M. 1988, \pasp, 100, 1134

\bibitem[Cassan et al.(2012)]{cassan12}
Cassan, A., Kubas, D., Beaulieu, J.-P., et al. 2012, \nat, 481, 167

\bibitem[Dominik(1998)]{dominik98}
Dominik, M. 1998, \aap, 329, 361

\bibitem[Dong et al.(2009)]{dong09}
Dong, S., Gould, A., Udalski, A., et al.\ 2009, \apj, 695, 970

\bibitem[Gaudi (2012)]{gaudi12}
Gaudi, B. S. 2012, \araa, 50, 411

\bibitem[Gould(1992)]{gould92a}
Gould, A. 1992, \apj, 392, 442

\bibitem[Gould(2004)]{gould04}
Gould, A. 2004, \apj, 606, 319

\bibitem[Gould(2008)]{gould08}
Gould, A. 2008, \apj, 681, 1593

\bibitem[Gould \& Loeb(1992)]{gould92b}
Gould, A., \& Loeb, A. 1992 \apj, 306, 104

\bibitem[Gould et al.(2010)]{gould10}
Gould, A., Dong, S., Gaudi, B. S., et al. 2010, \apj, 720, 1073

\bibitem[Gould et al.(2006)]{gould06}
Gould, A., Udalski, A., An, D. 2006, \apjl, 644, L37

\bibitem[Gould et al.(2014)]{gould14}
Gould, A., Udalski, A., Shin, I.-G., et al. 2014, Sci, 345, 46


\bibitem[Han (2006)]{han06}
Han, C. 2006, \apj, 638, 1080

\bibitem[Han et al.(2013)]{han13}
Han, C., Udalski, A., Choi, J.-Y., et al. 2013, \apjl, 762, L28

\bibitem[Han et al.(2016)]{han16}
Han, C., Udalski, A., Gould, A., et al. 2016, \apj, 828, 53

\bibitem[Henderson et al.(2016)]{henderson16}
Henderson, C. B., Poleski, R., Penny, M., et al. 2016, \pasp, 128, 124401

\bibitem[Hwang et al.(2018)]{hwang18}
Hwang, K.-H., Udalski, A., Shvartzvald, Y., et al. 2018, \aj, 155, 20

\bibitem[Ida \& Lin(2004)]{ida04}
Ida, S., \& Lin, D. N. C. 2004, \apj, 616, 567

\bibitem[Jung et al.(2013)]{jung13}
Jung, Y. K., Han, C., Gould, A., \& Maoz, D. 2013, \apjl, 768, L7J

\bibitem[Jung et al.(2018)]{jung18}
Jung, Y. K., Udalski, A., Gould, A., et al. 2018, \aj, 155, 219

\bibitem[Jung et al.(2015)]{jung15}
Jung, Y. K., Udalski, A., Sumi, T., et al. 2015, \apj, 798, 123

\bibitem[Kayser et al.(1986)]{kayser86}
Kayser, R., Refsdal S., \& Stabell, R. 1986, \aap, 166, 36

\bibitem[Kervella et al.(2004)]{kervella04}
Kervella P., Th\'{e}venin F., Di Folco E., S\'{e}gransan D., 2004, \aap, 426, 297

\bibitem[Kim et al.(2016)]{kim16}
Kim, S.-L., Lee, C.-U., Park, B.-G., et al. 2016, JKAS, 49, 37

\bibitem[Kim et al.(2018)]{kim18}
Kim, D.-J., Kim, H.-W., Hwang, K.-H., et al. 2018, \aj, 155, 76


\bibitem[Koshimoto et al.(2017)]{koshimoto17}
Koshimoto, N., Udalski, A., Beaulieu, J. P., et al. 2017, \aj, 153, 1


\bibitem[Mr{\'o}z et al.(2017a)]{mroz17a}
Mr{\'o}z, P., Han, C., Udalski, A., et al. 2017a, \aj, 153, 143

\bibitem[Mr{\'o}z et al.(2018)]{mroz18}
Mr{\'o}z, P., Ryu, Y.-H., Skowron, J., et al. 2018, \aj, 155, 121

\bibitem[Mr{\'o}z et al.(2017b)]{mroz17b}
Mr{\'o}z, P., Udalski, A., Skowron, J. et al. 2017b, \nat, 548, 183


\bibitem[Muraki et al.(2011)]{muraki11}
Muraki, Y., Han, C., Bennett, D. P., et al. 2011, \apj, 741, 22


\bibitem[Nagakane et al.(2017)]{nagakane17}
Nagakane, M., Sumi, T., Koshimoto, N., et al. 2017, \aj, 154, 35


\bibitem[Nataf et al.(2013)]{nataf13}
Nataf, D. M., Gould, A., Fouqu\'{e}, P., et al. 2013, \apj, 769, 88

\bibitem[Paczy\'{n}ski(1986)]{paczynski86}
Paczy\'{n}ski, B. 1986, \apj, 304, 1

\bibitem[Pecaut \& Mamajek(2013)]{pecaut13}
Pecaut, M. J., \& Mamajek, E. E. 2013, \apjs, 208, 9

\bibitem[Pejcha \& Heyrovsk\'{y}(2009)]{pejcha09}
Pejcha, O., \& Heyrovsk\'{y}, D. 2009, \apj, 690, 1772

\bibitem[Pepe et al.(2011)]{pepe11}
Pepe, F., Lovis, C., \& S\'{e}gransan, D. 2011, \aap, 534, A58


\bibitem[Poleski et al.(2014)]{poleski14}
Poleski, R., Skowron, J., Udalski, A., et al. 2014, \apj, 795, 42

\bibitem[Schneider \& Weiss(1987)]{schneider87}
Schneider, P., \& Weiss, A. 1987, \aap, 171, 49


\bibitem[Shvartzvald et al.(2017)]{shvartzvald17}
Shvartzvald, Y., Yee, J. C., Calchi Novati, S., et al. 2017, \apjl, 840, L3


\bibitem[Skowron et al.(2011)]{skowron11}
Skowron, J., Udalski, A., Gould, A., et al. 2011, \apj, 738, 87


\bibitem[Skowron et al.(2016)]{skowron16}
Skowron, J., Udalski, A., Poleski, R., et al. 2016, \apj, 820, 4


\bibitem[Street et al.(2016)]{street16}
Street, R. A., Udalski, A., Calchi Novati, S., et al. 2016, \apj, 819, 93

\bibitem[Sumi et al.(2010)]{sumi10}
Sumi, T., Bennett, D. P., Bond, I. A., et al. 2010, \apj, 710, 1641

\bibitem[Sumi et al.(2011)]{sumi11}
Sumi, T., Kamiya, K., Bennett, D. P., et al. 2011, \nat, 473, 349

\bibitem[Suzuki et al.(2016)]{suzuki16}
Suzuki, D., Bennett, D. P., Sumi, T., et al. 2016, \apj, 833, 145

\bibitem[Szyma\'{n}ski et al.(2011)]{szymanski11}
Szyma\'{n}ski, M. K., Udalski, A., Soszy\'{n}ski, I., et al. 2011, AcA, 61, 83

\bibitem[Tenenbaum et al.(2014)]{tenenbaum14}
Tenenbaum, P., Jenkins, J. M., Seader, S., et al. 2014, \apjs, 211, 6

\bibitem[Udalski et al.(2018)]{udalski18}
Udalski, A., Ryu, Y.-H., Sajadian, S., et al. 2018, AcA, 68, 1

\bibitem[Yoo et al.(2004)]{yoo04}
Yoo, J., DePoy, D. L., Gal-Yam, A., et al. 2004, \apj, 603, 139

\bibitem[Zang et al.(2018)]{zang18}
Zang, W., Penny, M. T., Zhu, W., et al. 2018, \pasp, submitted, arXiv:1803.09184

\bibitem[Zhu et al.(2017)]{zhu17}
Zhu, W., Udalski, A., Calchi Novati, S., et al. 2017, \aj, 154, 210


\end{thebibliography}
\end{document}